\documentclass[journal,onecolumn]{IEEEtran}
\IEEEoverridecommandlockouts
\usepackage{cite}
\usepackage{amsmath,amssymb,amsfonts}
\usepackage{algorithmic}
\usepackage{graphicx}

\usepackage{subcaption}

\usepackage{textcomp}
\usepackage{xcolor}
\def\BibTeX{{\rm B\kern-.05em{\sc i\kern-.025em b}\kern-.08em
    T\kern-.1667em\lower.7ex\hbox{E}\kern-.125emX}}
\begin{document}

\title{A Deep Recurrent Q Network towards Self-adapting Distributed Microservices architecture}

 \author{\IEEEauthorblockN{Basel Magableh}\\
\IEEEauthorblockA{\textit{School of Computer Science, } \\
\textit{Dublin Institute of Technology,}\\
\textit{Technological University}\\
Dublin, Ireland \\
basel.magableh@dit.ie}}
\maketitle

\begin{abstract}
One desired aspect of microservices architecture is the ability to self-adapt its own architecture and behaviour in response to changes in the operational environment. To achieve the desired high levels of self-adaptability, this research implements the distributed microservices’ architectures model, as informed by the MAPE-K model. The proposed architecture employs a multi adaptation agents supported by a centralised controller, that can observe the environment and execute a suitable adaptation action. The adaptation planning is managed by a deep recurrent Q-network (DRQN). It is argued that such integration between DRQN and MDP agents in a MAPE-K model offers distributed microservice architecture with self-adaptability and high levels of availability and scalability. Integrating DRQN into the adaptation process improves the effectiveness of the adaptation and reduces any adaptation risks, including resources over-provisioning and thrashing. The performance of DRQN is evaluated against deep Q-learning and policy gradient algorithms including: i) deep q-network (DQN), ii) dulling deep Q-network (DDQN), iii) a policy gradient neural network (PGNN), and iv) deep deterministic policy gradient (DDPG). The DRQN implementation in this paper manages to outperform the above mentioned algorithms in terms of total reward, less adaptation time, lower error rates, plus faster convergence and training times. We strongly believe that DRQN is more suitable for driving the adaptation in distributed services-oriented architecture and offers better performance than other dynamic decision-making algorithms.   
\end{abstract}


\section{Introduction}

 Self-adaptability refers to the ability of service oriented architecture (SOA) to modify its own structure and behaviour in response to changes in the operating environment \cite{Oreizy:1999p3722}. High levels of self-adaptability present the challenges of self-organising, self-tuning, and self-healing the architecture against an interruption. Moreover, because of the services' pervasiveness, and in order to make any adaptation strategy effective and successful, adaptation actions must be considered in conjunction with services availability, dependability \footnote{Software dependability refers to the degree to which a software system or component is operational and accessible when required for use.}, and reliability by providing an intelligent selection of the adaptation actions. So that the performed action meets the adaptation goals, objectives, and the desired architecture quality attributes \cite{Cheng:2009p3763,bailey2011self,Barbacci:2010p4077}. Thus, adaptation planning requires mechanisms that are able to learn how to choose adaptation actions from continuous actions space. The adaptation strategy must able to optimise the adaptation actions in order to guarantee that the architecture will reach the adaptation objectives \cite{van2012reinforcement}. On the other hand,  reinforcement learning (RL) provides software agents with the possibility to learn a specific policy that can be used to take decisions among a set of actions by maximising  the cumulative rewards yielded from executing a specific action \cite{silver2016mastering}
 
 Assuming that the SOA has a finite number of states and actions, and the SOA design confirms to the the MAPE-K model (monitor, analyse, plan, execute over a shared knowledge) \cite{computing2006architectural}. The reinforcement learning (RL) algorithm can be used for planning the adaptation and learning the rewards to be gained from performing each action. However, supporting the adaptation planning requires i) a sequential decision making in a continuous domain of states and actions, and ii) a mechanism to calculate the result reward of the adaptation. Unfortunately, the reward will be unknown and delayed until the adaptation action is completed and the architecture enters its new state.

 %In this context, the adaptation execution follows Markov Decision Process (MDP). MDP defined as set of states $s \in S$ and actions $a \in A$.  The transition model from  state $s$ to state $\tilde{s}$ is defined as a function $T(s,a,\tilde{s})$ and the reward of this action in the new state $\tilde{s}$ is defined by $R(s,a,\tilde{s})$, which return a real value every time the system moves from one state to another. Adaptation agent has no idea what the transition probabilities are! It does not know $T(s, a, \tilde{s})$, and it does not know what the rewards are going to be either (it does not know $R(s, a, \tilde{s})$) once it moves from one state to another. 

This research implements SOA as a distributed microservices architecture running in Docker swarm \footnote{https://docs.docker.com/engine/swarm/} as distributed cluster of workers and managers nodes. Docker swarm enforces the cluster to have one single leader following the implementation of Raft consensus algorithm \cite{ongaro2015raft}. Consequently, we consider the problem of sequential decision making in a continuous domain with delayed reward signals. The problem of delayed noisy rewards can be found in distributed microservices architecture. The full problem requires an algorithm to learn how to choose an action from infinitely large action space, in order to optimise a noisy delayed cumulative reward in a large state space, where the outcome of a single action can be stochastic \footnote{Stochastic: having a random probability distribution or pattern that may be analysed statistically but may not be predicted precisely}, because each node in the cluster could have different reward value as result of the selected action. The challenges of distributed microservices architecture are: a) the context model is unknown at runtime, b) each node might have different values of the observation space as they are naturally distributed, and c) each node could calculates different reward for each pair of state-action.

This paper studies the problem of supporting dynamic adaptation by multi agents centralised reinforcement learning, that can yield the highest rewards during dynamic adaptation process. In reinforcement learning this problem is solved using: i) policy gradient, ii) Q value, and iii) Q learning \cite{van2012reinforcement}; however, it is unknown for us which approach is more suited to the domains of self adaptive distributed service oriented architecture with a centralised controller. 

 The objectives of this research are: a) to provide a test bed of self-adapting distributed microservices architecture, that can be used by other researchers to experiment with various types of adaptation techniques. b) to evaluate which reinforcement learning algorithm is more suitable to support dynamic adaptation in microservices cluster. c) to propose a service stack that can be used to implement a distributed microservices architecture that confirms to the MAPE-K model. d) to propose adaptation agents that i) implement a Markov decision process (MDP) environment \cite{bellman1957markovian}, ii) are able to collect observations about the environment and iii) can execute adaptation actions. e) to evaluate the effectiveness  of reinforcement learning algorithms in dynamic selection of adaptation actions in distributed microservices architecture. 

In this research, five reinforcement learning algorithms are implemented to drive the adaptation process in self-adaptive  architecture. This paper will evaluate the following Q-learning algorithms: 1) deep Q networks (DQN) \cite{van2012reinforcement}, 2) duelling deep Q networks (DDQN) \cite{wang2015dueling}, 3) deep recurrent Q networks (DRQN) \cite{hausknecht2015deep}), as well as policy gradient algorithms: 4) policy gradient neural network (PGNN) \cite{sutton2000policy}, 5) deep deterministic policy Gradient (DDPG) \cite{silver2014deterministic}). Also, this paper will explain in detail the model that was used for building the microservices cluster, and the implementation of the experiment.  

%we will apply Q learning and policy gradient to provide the architecture with a possibility to learn adaptation policy that can be used to take decision dynamically at runtime among a set of adaptation actions.
 %There are two popular approaches in deep RL algorithms: Deep Q Networks (DQN) and policy gradient

This paper is structured as follows:Section \ref{sec:related-work} provides an overview of self-adaptation techniques and surveys the approaches used for context sensing, adaptation planning and reinforcement learning. Section \ref{sec:modelling} presents a model of a distributed microservices architecture that can continuously observe and adapt the architecture. The proposed method of adaptation planning and execution is discussed in Section \ref{sec:mapk}. Section \ref{sec:dqn} discusses different approaches for implementing reinforcement Learning algorithms. The calculation of the reward function is explained in section \ref{sec:reward}. \ref{sec:setup} addresses the implementation of this model and give more details about the architecture of the implementation of the five reinforcement learning algorithms. Section \ref{sec:evaluation} evaluates the effectiveness of the five reinforcement learning algorithms (DQN, DDQN, DRQN, PGNN, and DDPG) in adaptation planning and execution. 
Section \ref{sec:Conclusion} summarises this research, highlighting its contribution and setting out potential future work.
\section{Related work}
\label{sec:related-work}
\subsection{Self-adaptive service oriented architecture}
Self-adaptive architecture is characterised by a number of properties best referred to as autonomic \cite{jelasityself}. These properties are grouped under the `self-* properties' heading; they include self-organisation, self-healing, self-optimisation and self-protection  \cite{horn:2001p3735}. Self-adapting architecture refers to the capability of discovering, diagnosing and reacting to disruptions. It can also anticipate potential problems and, accordingly, take suitable actions to prevent a system failure \cite{horn:2001p3735}. Self-adapting aspects of service oriented architecture require a decision-making strategy that can work in real-time. This is essential for the architecture to reason about their own state and its surrounding environment in a closed control loop model and act appropriately \cite{Cheng:2008p3708}.  
Typically, a self-adapting system follows MAPE-K model (monitor-analyse-plan-execute over a shared knowledge). an efficient self-adaptive system should implements MAPE-K model which will include: a) gathering of data related to the surrounding context (context Sensing); b) context observation and detection; c) dynamic decision making; d) adaptation execution to achieve the adaptation objectives defined as QoS; e) verification and validation of the applied adaptation actions in terms of the ability of the taken action to meet the adaptation objectives and meet the desired QoS.

 However, Many approaches are used for achieving high levels of self-adaptability though context sensing; a model that involves context collection, observation and detection of contextual changes in the operational environment \cite{Strang:2004p3770}. Also, the ability of the system to dynamically adjust its behaviour can be achieved using parameter-tuning \cite{Cheng:2009p3902}, component-based composition \cite{MariusMikalsen:2005ur}, or middleware-based approaches \cite{CheungFooWo:2007p1692}. Another important aspect of self-adaptive system is related to its ability to validate and verify the adaptation action at runtime based on game theory \cite{Wei:2016ge}, utility theory as in \cite{Menasce:2007vq,KonstantinosKakousis:2008ub}, or a model driven approach as in \cite{Sama:2008p3765}.

\textit{Context information} refers to any information that is computationally accessible and upon which behavioural variations depend \cite{Hirschfeld:2008p1620}. \textit{Context observation and detection approaches} are used to detect abnormal behaviour within the architecture at run-time. Related work in context modelling, context detection and engineering self-adaptive software systems are discussed in \cite{Salehie:2009p3693,Cheng:2008p3708,RogeriodeLemos:2011tj,Strang:2004p3770}.  
In \textit{dynamic decision making} and \textit{context reasoning} the architecture should be able to monitor and detect normal/abnormal behaviour by continuously monitoring the contextual changes found in the operational environment. 

In distributed services oriented architecture, the performance of the cluster's nodes could fluctuate around the demand to accommodate scalability, orchestration and load balancing issued by cluster's leader.  
This behaviour requires a model that is able to detect anomalies in real-time and which can generate a high rate of accuracy in detecting any anomalies and a low rate of false alarms. In addition, there will be a set of variations that can be used by the system to adapt the changes in its operational environment. This adaptation requires dynamic decision-making process that can maximises the cumulative rewards gained from executing a specific action. Such problem is solved by reinforcement learning algorithms \cite{van2012reinforcement} as discussed in the following section. %Each node in the cluster will be able to observe the reward executed by the leader node in the cluster. This research focus on studying self-adaptation in centralised distributed microservices architecture, supported by reinforcement learning algorithms. Decentralised multi-agents reinforcement learning algorithms are out of the scope of this paper. In the following section, we will focus on describing RL approaches to support dynamic adaptation and the selection of adaptation actions. 

\subsection{Reinforcement learning in continuous state and action spaces}

\subsubsection{Q-value:}
Bellman \cite{bellman1957markovian} found an algorithm to estimate the optimal state-action values called Q-values. The Q-value of a state-action pair is noted by $Q(s,a)$. The $Q(s,a)$ refers to the sum of discounted future rewards that the action expects to reach in a state $s$ after selecting the action $\alpha$. Q-value estimation is not applicable in environments with large sets of states and actions. Alternatively, for large state and action spaces two popular approaches were proposed i) Q-learning and ii) policy gradient \cite{van2012reinforcement}. 
\subsubsection{Q-learning:}
in Q-learning, neural network are used to estimate the Q-value by defining approximation function and train the model in deep Q-networks; an approach is called deep Q-learning. Deep Q-learning is a multi-layered neural networks that for a given state $s$ outputs a vector of action values using Markov's decision process \cite{bellman1957markovian} which uses approximation function to estimate the Q-value $Q(s, a)$. The goal of DQN is to learn a state-action value function (Q), which is given by the deep networks, by minimising temporal-difference errors \cite{mnih2013playing}. 

Several efforts were made to employ neural networks in the implementation of RL algorithms \cite{van2012reinforcement}. The idea is to use the DQN to identify the mathematical relationship between the input data and to identify the maximum reward function of finding the output. Such efforts can be found in \cite{mnih2013playing} where RL and NN were used to play Atari games or Go games as in \cite{sutton1998introduction}. Also, several methods were proposed to solve  computer vision problems \cite{yoo2017action}, such as object localisation \cite{caicedo2015active} or action recognition \cite{jayaraman2016look}, by employing the deep RL algorithms.
Based on the DQN algorithm \cite{van2012reinforcement}, various neural networks such as double DQN \cite{van2016deep} and duelling DQN \cite{wang2015dueling} were proposed to improve performance and keep stability. Also, recurrent neural networks (RNN) were employed to overcome the problem of partial observability as proposed  in \cite{hausknecht2015deep}. A combination between Long Short Term Memory (LSTM) \cite{hochreiter1997long} and RNN to enhance the knowledge of the agents about the observable environment, or what so called Partially Observable Markov Decision Processes (POMDPs). In POMDPs environment the agent has only a partial view of the full environment's state space and action space, which leads to poor performance of the Q-learning algorithms. The use of deep recurrent Q-learning combined with LSTM enables the agent to build more knowledge about the transformation model and the cumulative rewards yielded from moving between states. The use of deep recurrent Q-learning network was found to be a very promising approach for planning the adaptation in microservices architecture.

RNN offers many features that suit the nature of self-adapting architecture. Self-adaptive service oriented architecture can be described as POMDPs environment at state $s$ the adaptation agent has no full view of the unforeseen context changes in the next future state. Also, the DQN uses forward-propagation to estimate the Q-value, then DQN checks the error between the predicted Q-value and the true actual yielded Q-value from each state-action pair, which leads to poor performance in real world applications and slow convergence \cite{hausknecht2015deep}. 

On the other hand, the DRQN by employing RNN and LSTM it is able to i) reveal the patterns between the collected observations and the cumulative yielded rewards for each pair of state-action. ii) DRQN has the ability to perform backpropagation through time, to find the derivatives of the error with respect to the calculated weights of the observations, which enables the DRQN to improve the gradient descent, and subsequently minimizes the loss and error rate of the predicted Q-value. iii) LSTM enables the DRQN to possess a memory of all the yielded rewards from each pair of state-action pairs, so that the DRQN will be able to identify which action sequences return the maximum reward and consequently DRQN reaches a terminal state in less time than DQN. iv) the problematic issues of credit assignment and temporal-difference are solved by the ability of the DRQN to feed the current state output as an input for the next state combined with the collected observations which improve its accuracy and significantly reduces the required training time.

%In self-adaptive service oriented architecture, the context changes are unforeseen and has high level of uncertainty, which results in high rate of changes and causes the problem of temporal difference. The RNN is suitable to capture the temporal difference of the context changes as it can i) recognizes patterns in sequences of data streams. ii) Backpropagation Through Time, In neural networks, you basically do Forward-Propagation to get the output of your model and check if this output is correct or incorrect, to get the error. Now you do Backward-Propagation, which is nothing but going backwards through your neural network to find the partial derivatives of the error with respect to the weights, which enables you to subtract this value from the weights. Those derivatives are then used by Gradient Descent, which minimizes the loss and error rate. iii) An RNN can simultaneously take a sequence of inputs and produce a sequence of outputs. iv) LSTM’s enable RNN’s to remember their inputs over a long period of time. This is because LSTM’s contain their information in a memory, that is much like the memory of a computer because the LSTM can read, write and delete information from its memory. The problematic issues of vanishing gradients is solved through LSTM because it keeps the gradients steep enough and therefore the training relatively short and the accuracy high.

\subsubsection{Policy gradient}
On the other hand, the idea of a policy gradient algorithm is to update the adaptation policy with gradient ascent on the cumulative expected Q-value. So, the algorithm learns directly the policy function $\Pi(\alpha, s)$ without worrying about the Q-value \cite{williams1992simple}. If the gradient is known, the algorithm will update the policy parameters with probability that the agent will take action $\alpha$ in state $s$. However, we use Deep Neural Network to find the policy $\pi(s,a,\theta)$, given the state $S$ with parameters $\theta$. The network takes state as an input and produces the probability distribution of actions. Such approach of policy gradient can be found in policy gradient (PG) \cite{sutton2000policy}, Deep Deterministic Policy Gradient (DDPG) \cite{silver2014deterministic}). 

Policy gradient methods directly learn the policy by optimizing the deep policy networks with respect to the expected future reward using gradient descent. Williams et al. \cite{williams1992simple} proposed REINFORCE algorithm which simply uses the immediate reward to estimate the value of the policy. Silver et al. \cite{silver2014deterministic} proposed a deterministic algorithm to improve the performance and effectiveness of the policy gradient in high-dimensional action space. Silver et al. \cite{silver2014deterministic} showed that pre-training the policy networks with supervised learning before employing policy gradient can improve the performance of PG algorithms.  
\subsection{Reinforcement Learning in Self-adaptive Architecture}
The employment of reinforcement learning for dynamic action selection is not new to the domain of self-adaptive architecture. Several approaches were proposed including parameter tuning as can be found in \cite{ dowling2004self,salehie2005policy,tesauro2007reinforcement, amoui2008adaptive}. Model-based adaptation that involve architecture-based configurations was proposed in \cite{kim2009reinforcement}. Kim and Park \cite{kim2009reinforcement} demonstrated the use of Q-learning to perform architectural changes in a simulated robot. This early work of Kim and Park uses a Q-table for all state-action pairs that can be used to calculate the Q-value and dynamically the algorithm selects an adaptation action. 

However, the above mentioned approaches faces many limitations including that they attempt to apply reinforcement learning in discreet state and action space limited to a specific domain, which in turn limits the generalisation of those approaches as presented in \cite{tesauro2007reinforcement, salehie2005policy}. More recently, reinforcement learning was proposed as a method to adjust resource allocation in cloud infrastructure in \cite{dutreilh2011using, jamshidi2016fuzzy}. Finally, Tong et al. \cite{wu2018using} employs reinforcement learning to handle context uncertainty in self-adaptive systems. 

Handling the unforeseen contextual changes in dynamic software systems is a complex problem and a changeling task that exceeds the assumption of having a discrete state and action spaces as specified by \cite{wu2018using}. Uncertainty of context changes requires a mechanism to handle continuous state and/or action spaces; a situation which requires a mechanism to represent the context model at runtime to reflect the recent changes in the operational environments. In addition to that, context uncertainty in distributed self-adaptive systems requires a multi decentralised adaptation agents that could possibly adapt to changes in distributed system. Decentralised multi-agent deep reinforcement learning (DMARL) is an ongoing research filed as described in \cite{busoniu2008comprehensive, marinescu2014decentralised,lowe2017multi,zhang2018fully}.
Learning in decentralised multi-agent environment is fundamentally more challenging than the employment of single agent. DMARL faces serious problem like having non-stationary state, high dimensionality of the observation space, multi-agent credit assignment, robustness, and scalability \cite{hernandez2018multiagent}, so employing DMARL in self-adaptive architecture is a  challenging task, that could be investigated in future work.

Mnih et al. \cite{mnih2016asynchronous} proposes asynchronous advantage actor-critic (A3C) algorithm to overcome the issue of multi-agent credit assignment. Multiple workers are allowed to estimate the Q-value for each state-action pair and single global network decided which policy is more suitable for the state. It is a combination between policy gradient and Q-learning that allows parallel deep Q-networks to work asynchronously. A2C is a synchronous, deterministic version of A3C \cite{mnih2016asynchronous}. This research implements DDPG a similar algorithm to A2C. A3C could be investigated in future, but the problem of A3C that some workers will estimate the Q-value based on an older version of the parameters $\theta$, this might add more complexity in top of the well known issue of eventual vs strong consistency found in distributed systems. So we decided to conduct our research in a centralised multi-agent microservices architecture, where all agents are able to observe the environment, and only the manager nodes will execute the selected action by RL algorithms. Keeping in mind, that RAFT consensus algorithm \cite{ongaro2015raft} is a central component of the implementation of Docker swarm cluster. So, the selected actions by the deep RL algorithms are a) subject to the voting process of RAFT consensus algorithm \cite{ongaro2015raft}. b) multiple agents are observing the environment, but the execution of the adaptation actions are limited to manager nodes, supported by one single leader at a time.

In Docker swarm \footnote{https://docs.docker.com/engine/swarm/} manager nodes are used to handle the  management of the cluster's tasks and maintaining cluster's state, scheduling the microservices, and provide access to the  microservices endpoints in global mode. Meaning, microservices running in worker nodes can be accessible via the managers' endpoints \cite{kiss2017micado}. The challenge of Docker swarm cluster is that the manager's quorum are stored in all manager nodes, but only the  elected leader makes all swarm management and orchestration decisions for the swarm. The quorum are used to store information about the cluster states, and the consistency of information is achieved through consensus via the Raft consensus algorithm \cite{ongaro2015raft}. Once the leader node dies unexpectedly, other managers can be elected to serve as a leader for the swarm, then the leader restores the cluster state. This particular implementation of Raft in Docker swarm enforces this research to use multi-agent with centralised controller (i.e. swarm leader), so that decentralised multi agent reinforcement learning is beyond the scope of this paper.

\section{Adaptation Strategy}
\label{sec:modelling}

This section focuses on describing a model that can continuously observe and monitor microservices architecture and be able to provide an adaptation action that can maintain the architecture's availability. At the same time, the architecture should be able to respond to unforeseen changes by selecting the best adaptation actions, that optimise the computational resources and achieve the adaptation goals.
 
The proposed model consisted of: 1) Service oriented architecture implemented as distributed microservices architecture. 2) Adaptation (MDP) multi-agents that observes the execution environment and executes the adaptation action. 3) an adaptation planning component implemented by using one of the Q-learning or policy gradient algorithms. The implementation details of the microservices used in this architecture are explained in Section \ref{sec:setup}. To understand the proposed model of self-adaptive microservices architecture, it is necessary to understand the functionality of the adaptation agents as described in the following section.
 
\subsection{Adaptation MDP Agents}
\label{sec:mapk}
The adaptation agents follows Markov decision process (MDP). MDP can be defined as a set of states $s \in S$ and actions $a \in A$. The transition model from  state $s$ to state $\tilde{s}$ is defined as a function $T(s,a,\tilde{s})$ and the reward of this action in the new state $\tilde{s}$ is defined by $R(s,a,\tilde{s})$, which returns a real value every time the system moves from one state to another. In service oriented architecture the set of possible adaptation actions is identified a) based on the observation of the current state of the architecture and b) by proposing a suitable action that can adapt to the contextual changes. The adaptation requires the identification of a set of sequential actions to adapt to changes by executing an adaptation action that enables the architecture to reach the desired objectives or quality of services. As a result the adaptation agent will be rewarded with positive value once it reach the adaptation objectives (i.e. service convergence) or negative reward value for every failed adaptation action. Having a noisy delayed cumulative reward value prevents the agents from identifying the best action to take. The problem of noisy reward is solved by estimating the Q-value using a reinforcement learning approach as discussed in the following section.

\subsection{Reinforcement Learning in Continuous State and Action Spaces}
\label{sec:dqn}
However, the adaptation agents with discrete adaptation actions has no idea what the transition probabilities are! It does not know $T(s, a, \tilde{s})$, and it does not know what the rewards are going to be either (it does not know $R(s, a, \tilde{s})$) once it moves from one state to another. The agents must experience each state and each transition at least once to know the rewards, and it must experience them multiple times if it is to have a reasonable estimate of the transition probabilities. Knowing the optimal state value is very useful to identify the best adaptation action(s). Bellman \cite{bellman1957markovian} found an algorithm to estimate the optimal state-action values called the Q-values. The Q-value of a state-action pair is noted by $Q(s,a)$. The $Q(s,a)$ refers to the sum of discounted future rewards that the adaptation action expects to reach in a state $s$ after selecting the adaptation action $\alpha$. The Q-value estimation is not applicable in environment with large set of states and actions. Alternatively, there are two more popular approaches: Q-learning and policy gradient algorithms. 

In Q-learning, neural network are used to estimate the Q-value by defining approximation function and train the model in deep neural network, this approach is called Deep Approximate learning (Deep Q-learning). Deep Q-learning is a multi-layered neural network that for a given state $s$ outputs a vector of action values and it uses approximation function to estimate the Q-value $Q(s, a)$. 

On the other hand, the idea of a policy gradient algorithm is to update the adaptation policy with gradient ascent on the cumulative expected Q-value. So, the algorithm learns directly the policy function $\Pi(\alpha, s)$ without worrying about the Q-value \cite{williams1992simple}. If the gradient is known, the algorithm will update the policy parameters with the probability that the agents will take action $\alpha$ in state $s$. However, we use deep neural network to to find the policy $\pi(s,a,\theta)$, given the state $s$ with parameters $\theta$. $\theta$ refers to a selected well crafted features of the observation space \cite{van2012reinforcement}. The network takes `state' as an input and produces the probability distribution of an action. The proposed adaptation planing will implement both approaches of Q-learning and policy gradient, then their effectiveness in driving the adaptation process will be evaluated in this study. To be able to understand the process of the adaptation it is vital to understand the reward function used in each approach and how it is calculated; issues that will be described in the following section. 

%This paper will evaluates: i) Deep Q Learning Network (DQN) \cite{van2012reinforcement}, ii) Duelling Deep Q Learning Network (DDQN) \cite{wang2015dueling,van2016deep}, Deep Recurrent Q Learning Network (DRQN) \cite{hausknecht2015deep}), iii) policy gradient neural networks (PGNN) \cite{sutton2000policy}, iv) Deep Deterministic Policy Gradient (DDPG) \cite{silver2014deterministic}). 

\begin{figure*}
\centering
\includegraphics[scale=0.48]{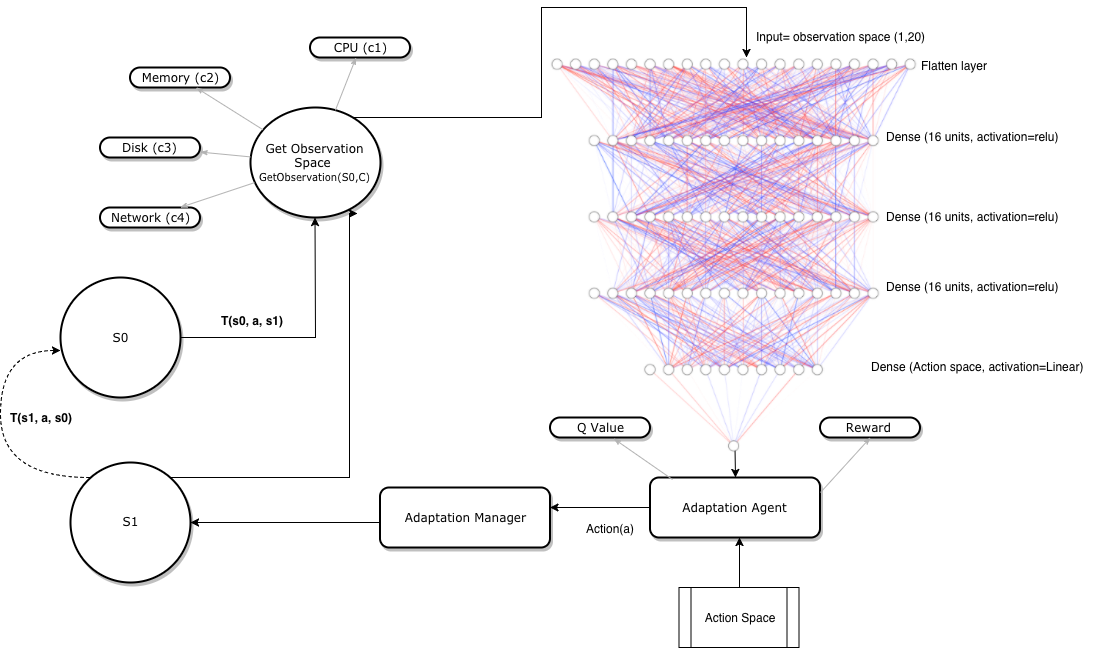}
\caption{Deep Q Network and transition model from S0 to S1}
\label{fig_Transiationmodel}
\end{figure*}

 \subsection{Reward Function}
 \label{sec:reward}
The adaptation agents will be using one deep Q-learning/policy gradient approaches for identifying the best adaptation action that can return the highest reward once it reaches the desired adaptation objective. For this aim, we need to define the reward function i.e. Q-value $Q(s,a)$. To achieve this, we need to look back at the state $s$ of the service architecture (see Figure \ref{fig_Transiationmodel}. At each state $s0$ in Figure  \ref{fig_Transiationmodel} there is a set of observations $c \in C$ measuring the matrices of operating environment such as: CPU, Memory, Disk I/O and Network  as shown in Figure \ref{fig_Transiationmodel}. 

The target is to provide the deep Q-learning or policy gradient algorithms with a scalable value that can be used to assign a weight $W(s,c)$ for all context values $c$ found in state $s$.  The value $w(s,c)$ is calculated using neural network integrated with deep q-learning/policy gradient algorithms. The deep Q-learning approach is a regression algorithm so we have used mean squared error as cost function and we minimise that loss during the training between the predicted/estimated Q-value and the actual Q-value. This makes the mean squared error a good criterion to estimate the performance of DQN, DDQN, and DRQN algorithms. On the other hand, the policy gradient approach is a classification algorithm so we use cross entropy to calculate the optimal policy. So the probability that a given sequence of actions occurs is equal to the probability that the corresponding sequence of states and actions occurs with the given policy. This makes the cross entropy measures the probability of executing a random selected action and it will return the loss between the calculated probability and the actual probability of the executed action. This makes the Softmax of cross entropy loss a good measurement of policy gradient algorithms.
 
The following section explains the experimental setup and the structure of neural networks used in the implementation of self-adaptive microservices architecture. 

\section{Experiment Setup} 
\label{sec:setup}

\subsection{Microservices Architecture implementation}
To validate the ideas presented in this paper, we developed a working prototype of service oriented architecture as microservices running in Docker swarm \footnote{https://docs.docker.com/engine/swarm/}. The cluster has only one single leader at a time supported by manager and worker nodes. Only the leader/manager node(s) executes the proposed adaptation actions, then all manager nodes will vote on the proposed action. The outcome of the vote is stored by the leader node's logs. The action will be executed after a successful result according to the mechanism explained by Raft consensus algorithm \cite{ongaro2015raft}. The leader node and all managers will be running the adaptation agents for observation and adaptation execution. The adaptation planning will be running at the leader node, where one single algorithm of Q-learning or policy gradient will be used to drive the adaptation and its evaluation measurements will be collected in separate experiment. 
 
To comply with the MAPE-K model, the microservices architecture implements the following services:  
\begin{itemize}
\item A time series metrics database for context collection implemented using  Prometheus framework \footnote{https://prometheus.io}
\item Nodes metrics used to collect metrics from all nodes in the cluster,
\item Docker containers metrics collector for collecting fine-grained metrics about all running containers in all nodes \footnote{https://github.com/google/cadvisor}
\item Alert and notification managers used to notify the adaptation manager about contextual changes \footnote{https://prometheus.io/docs/alerting/alertmanager/}.
\item Reverse proxy for routing traffic between all services in the cluster \footnote{https://caddyserver.com/docs/proxy}.
\item Time series analytics and visualisation dashboard for observing the behaviour of the microservices cluster by the end users \footnote{https://grafana.com}.
\item MDP adaptation agents deployed in all nodes, that can observes the microservices architecture and executes the selected adaptation action(s). 
\item Adaptation planning service running on the leader node, which allows the deep Q-networks to have a consistent value for all parameters. 
\end{itemize}
The Adaptation planning service is implemented via  (deep Q-learning network (DQN), Duelling Deep Q Learning Network (DDQN), Deep Recurrent Q Learning Network (DRQN), Policy Gradient Neural Network (PGNN), or Deep Deterministic Policy Gradient (DDPG)). The five algorithms were implemented  using the Keras-rl framework \cite{plappert2016kerasrl} and Tensorflow framework \footnote{https://www.tensorflow.org}. 

Each algorithm will be executed separately and the algorithm will collect the observation via the multi adaptation agents running in all nodes. The algorithm processes the observation and proposes an adaptation action according to their implementation of Q-learning or policy gradient. The selected action policy will be returned to the adaptation agent for execution. The leader node initiates the voting process over the selected action. Wining the vote will results of orchestrating the selected action to all nodes following the mechanism of Raft consensus algorithm \cite{ongaro2015raft}. The agents run the adaptation and return the new observation and the cumulative reward value yielded in the new state.

 To understand the implementation of the five reinforcement learning algorithm, we need to study the features of each algorithm as well as the structure of the deep Q-network used to estimate the Q-value/policy gradient. The implementation of the five algorithms are discussed in the following sections.  A full code of the adaptation agents, the five algorithms implementation and services stack used in this experiment can be found in \footnote{https://github.com/baselm/cmarl.git}

\subsection{Deep Q Learning Network (DQN)}
 
 \begin{figure}
 \centering
\includegraphics[scale=0.25]{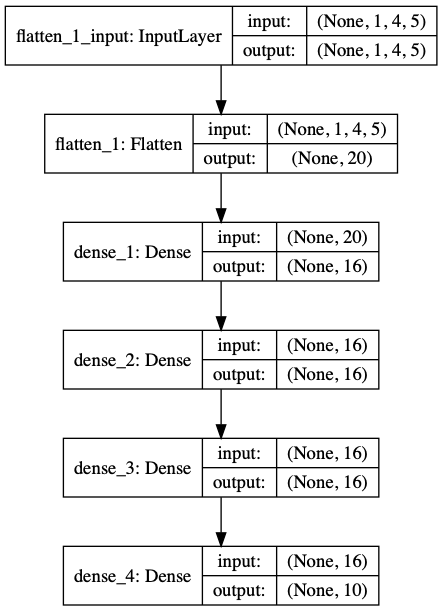}
\caption{Deep Q-learning Network}
\label{fig_dqn}
\end{figure}
  
 In our implementation of DQN it takes a state $s$ observation as input and outputs a Q-value estimate per action, which will be executed by the adaptation agent. The neural network architecture shown in Figure \ref{fig_dqn} consisted of an input layer equals the size of the observation space. In this experiment, the adaptation agent collects CPU usage, memory usage, disk space, network I/O.  The observation at each state is the input for the DQN first flatten layer (see Figure \ref{fig_dqn}). The observation space is multidimensional box, for this reason we used a flatten layer that flattens the observations into one dimensional input so they can be used by following dense layer. The second layer is dense hidden layer of 20 units. The RELU activation function \cite{nair2010rectified} is used to activate the output for the next dense layer. The last layer in the DQN in Figure \ref{fig_dqn} is  a dense layer with linear activation function. The output of dense-4 layer in Figure \ref{fig_dqn} equals the estimated Q-value of the associated action value of the action space defined by the adaptation agent. Our adaptation agent implements a discrete action space of 10 possible adaptation policy that can be used to: a) scale the microservices architecture horizontally or vertically, b) perform service composition, c) preserves the cluster state, and d) perform auto recovery. 

 However, the action space can be extended using requirements reflections \cite{Bencomo:2010p3675}, dynamic services compositions, or model-driven variability management as in \cite{Medvidovic:1996p3905,Paspallis:2009p3570}. Keeping in mind, that Docker swarm are supported particularly with the features of: i) services discovery, ii) services composition, iii) load balancing, and iv) orchestration. All features are managed under Raft consensus algorithm \cite{ongaro2015raft}, which makes extending the actions space an easy task by means of tailoring different variations of docker-compose files driven by business requirements or goal driven adaptation \cite{CheungFooWo:2007p1692}. 

\subsection{Duelling Deep Q Learning Network (DDQN)}

 \begin{figure}
 \centering
\includegraphics[scale=1.3]{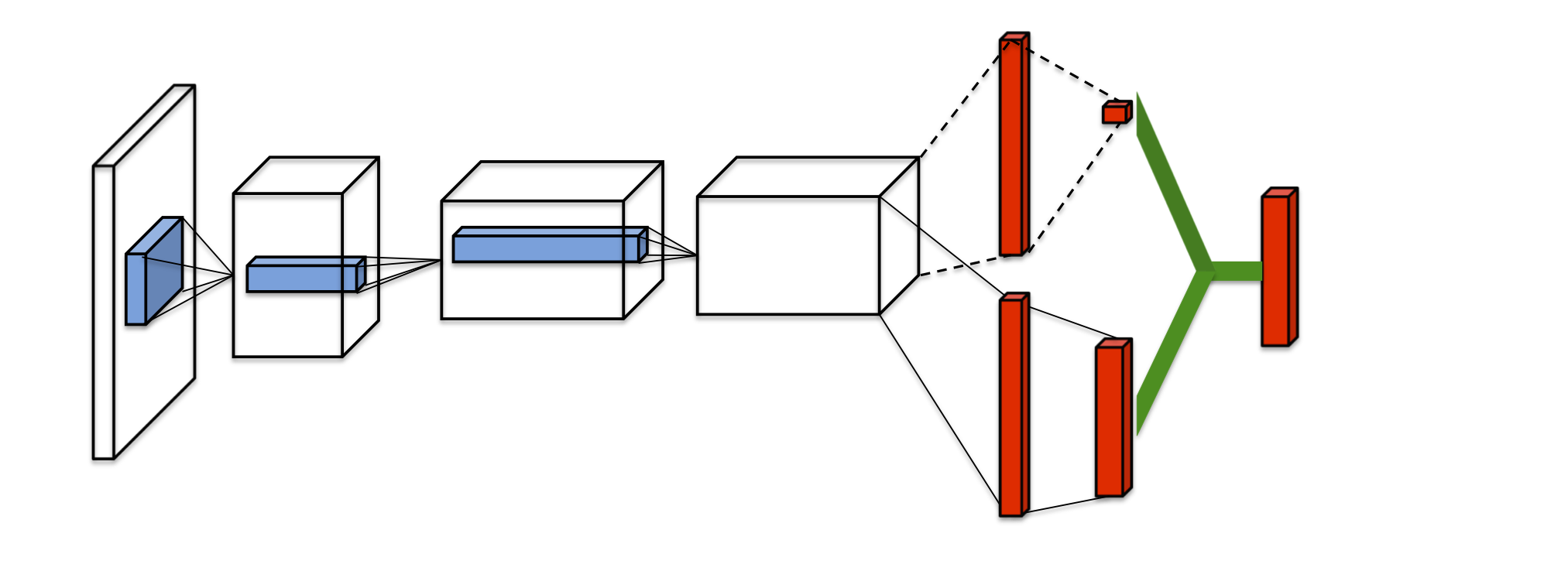}
\caption{Duelling Deep Q Learning Network Architecture (DDQN)}
\label{fig_ddqn2}
\end{figure}

The architecture of duelling deep Q-learning network (DDQN) is very similar to DQN illustrated in Figure \ref{fig_dqn}. The  modifications that DDQN algorithm offers are: a) it separates the representation of observation state from the representation of action space. b) DDQN calculates an estimation of the advantage and Q-value for each state-action pair separately in each layer, after which it will combines them back into a single Q-value at the final layer as shown in Figure \ref{fig_ddqn2}. The DDQN calculates the Q-value $Q(s,a)$ from the estimated Q-value $V(s)$ and the calculated advantage of state-action pair $A(s,\alpha)$, this results of calculating the Q-value as $Q(s,a) =V(s) + A(a)$. Wang et al. \cite{wang2015dueling} claims that DDQN speeds up the convergence better than DQN when the action space is very large. 
 
\subsection{Deep Recurrent Q Learning Network (DRQN)}
Considering the implementation of the microservices architecture proposed in this paper, you will find that the adaptation agent can observe the architecture but it has only a partial view of the full environment observations as it does not know the probabilities of $T(s, a, \tilde{s})$ and the $R(s, a, \tilde{s})$  in the new state $\tilde{s}$. This problem in MDP is referred to as partially observable Markov decision processes (POMDPs) \cite{monahan1982state}, which implies that the agents have no idea what is the transformation or the reward after executing a specific action. The 
solution of POMDPs in reinforcement learning involves the use of temporal difference algorithms \cite{van2012reinforcement} to provide the agent with better knowledge about the environment which leads to a better estimation of the Q-value. This can be achieved by stacking the observation instead of dealing with a single observation at a time. Then a memory is used to replay the last collected set of observation to the DQN, which gives the DQN more insights about the rate of change, bias, and gradient decent of the observations. This method works well in many real-world examples as demonstrated in \cite{mou2017deep,murad2017deep}. Microservices architecture often features incomplete and noisy states' information, which leads to partial observability and uncertainty of the context model. DQRN was proposed to overcome the problem of partial observability \cite{hausknecht2015deep}. In this paper, we considered combining the deep recurrent Q-network with Gated Recurrent Unit (GRU) cells \cite{cho2014properties} instead of using LSTM \cite{hausknecht2015deep}. The use of GRU cell offers RNNs better convergence and needs less training time in comparison to the use of LSTM cell in RNNs as demonstrated by \cite{chung2014empirical}. 

\begin{figure}
\centering
\includegraphics[scale=0.25]{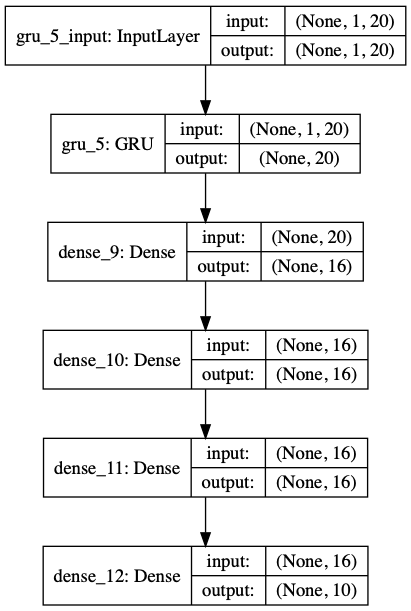}
\caption{Deep Recurrent Q Learning Network DRQN}
\label{fig_drqn}
\end{figure}

The implementation of DRQN is presented in Figure \ref{fig_drqn}. The DQN consisted of GRU cells that take the observation space as an input. The GRU is configured to return the last output in the output sequence as an input for the next dense layer. As shown in Figure \ref{fig_drqn}, the following layers are very similar to the architecture of DQN, where RELU used as an activation function until the output layer of the estimated Q-value is returned, a process which is associated with the  action policy to be executed by the adaptation agent.

%what the transition probabilities are! It does not know $T(s, a, \tilde{s})$, and it does not know what the rewards are going to be either (it does not know $R(s, a, \tilde{s})$) once it moves from one state to another. It must experience each state and each transition at least once to know the rewards, and it must experience them multiple times if it is to have a reasonable estimate of the transition probabilities.

\subsection{Policy Gradient with Deep Neural Network (PGNN)}
As was discussed above, the policy gradient optimises the policy parameters by following the gradient assent towards the highest reward. So the neural network policy will play several adaptation actions and compute the gradients at each step that would make the chosen action more likely without applying this adaptation in the architecture. If an action’s score is positive, it means that the action was good and the agent would apply the gradients computed earlier to make the action even more likely to be chosen in the future. However, if the score is negative, it means the action was bad and the agent would apply the opposite gradients to make this action slightly less likely in the future. The solution is simply to multiply each gradient vector by the corresponding action’s score. Finally, the PG calculates the mean of resulting gradient vectors and uses it to preform a gradient descent step. 

\begin{figure}
\centering
\includegraphics[scale=0.25]{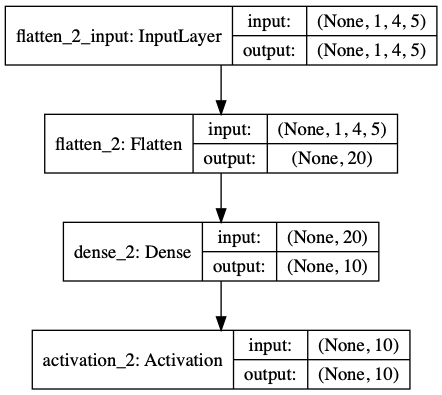}
\caption{Policy Gradient with Deep Neural Network (PGNN)}
\label{fig_pg}
\end{figure}

Our implementation of the policy gradient neural network (PGNN) algorithm involves the definition of the neural network shown in Figure \ref{fig_pg}. A simple neural network to predict the probability of taking an action using a cross entropy between the estimated probability and the target probability. The network shown in Figure \ref{fig_pg} consisted of a simple input layer and one single layer that outputs the action to be taken by the agent. The problem of having this simple neural network is that the rewards are sparse and delayed, so the network will only have the reward value to know whether an action was good or bad. In other words, the agent will fail to justify which action was responsible for positive/negative reward after executing a sequence of actions. To solve this problem, a common approach is to use a discount rate to evaluate an action based on the sum of all cumulative rewards that come after it by applying a discount rate $r$ at each step, then this reward will be normalised across the many iterations of the execution. 

 %If an action’s score is positive, it means that the action was good and you want to apply the gradients computed earlier to make the action even more likely to be chosen in the future. However, if the score is negative, it means the action was bad and you want to apply the opposite gradients to make this action slightly less likely in the future. The solution is simply to multiply each gradient vector by the corresponding action’s score.
\subsection{Deep Deterministic Policy Gradient (DDPG)}
The problem of having an environment with continuous state space or continuous action space is a challenging task for reinforcement learning. Having a continuous action space requires an algorithm that could provide better estimations of the best action to take, considering the partial observation problem found in the POMDPs environment. The deep deterministic policy gradient (DDPG) algorithm was proposed in \cite{lillicrap2015continuous} to overcome the issue of continuous action space. The idea of the DDPG algorithm is to have two neural networks, one acting as the critic and the other acting as the actor. The critic DQN is used to estimate the Q-value of a state-action pair. The actor is a policy gradient neural network that controls how the agent is behaving; judgement is based on the estimation it received from the critic network. Lillicrap at al. \cite{lillicrap2015continuous} claim that having a hybrid approach between Q-learning and policy gradient would overcome the problem of handling continuous actions space and the partial observability of the environment. 

\begin{figure*}[t!]
\centering
    \begin{subfigure}[b]{0.2\textwidth}
        \centering

        \includegraphics[scale=0.2]{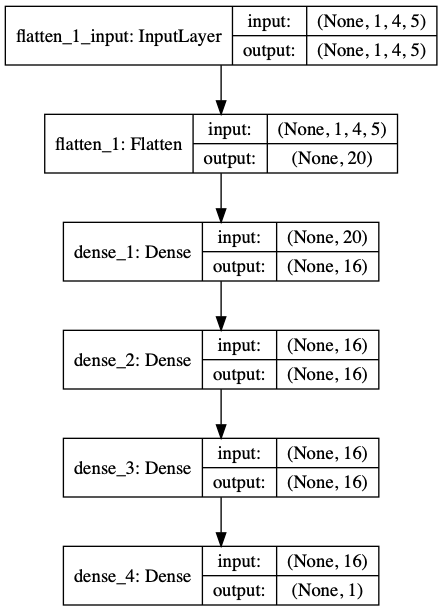}

        \caption{Actor Network\label{fig_actor}}
    \end{subfigure}
    \begin{subfigure}[b]{0.2\textwidth}
        \centering

        \includegraphics[scale=0.2]{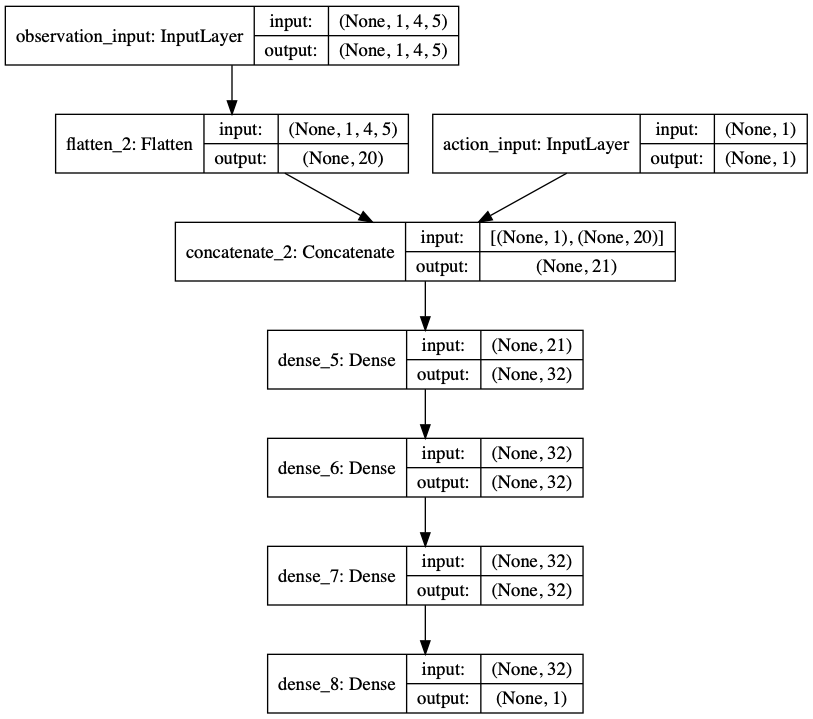}
        
        \caption{Critic Network\label{fig_critic}}
    \end{subfigure}
\caption{Deep Deterministic Policy Gradient (DDPG) }
\label{fig_DDPG}
\end{figure*}

The architecture of DDPG is shown in Figure \ref{fig_dqn}, the actor network is DQN similar to the architecture in Figure \ref{fig_dqn}. The first layer is an input flatten the observation space followed by three dense layers output the probability of an action. This output rather being sent to the agent is feed into the critic network. The critic network in Figure \ref{fig_critic} concatenates the output of the actor network and the observation space as a single input, then it feeds them into three consecutive dense layers, so it can identify the best probability for the state-action pair. We implement the critic network to use the Ornstein Uhlenbeck process for policy approximation as it offers better performance over epsilon-greedy approximation as stated in \cite{daprato1995ornstein}. 

 \section{Results}
 \label{sec:evaluation}
 Five different experiments were executed to evaluate the effectiveness of using the five algorithms in dynamic adaptation. From each experiment the following metrics are used to measure the algorithms' performance including: a) mean of the Q-value, b) total reward yielded per single episode, c) the adaptation time in seconds, which measures the time needed for each algorithm to complete action selection and the total number of steps taken by the algorithm at each episode to reach a terminal state. This measurement is very important to observe as some algorithms will become trapped in a state continuously believing it yields the maximum reward, which extended the adaptation time and delayed the transition to an optimal sate. d) the mean absolute error (MAE), which captures the error rate between the predicted Q-value and the actual Q-value obtained after executing the actions. e) number of steps per episode needed to reach a terminal result, which means that the adaptation agent manages to achieve full convergence of all services and cluster nodes.

 \begin{figure}
\centering
\includegraphics[scale=0.4]{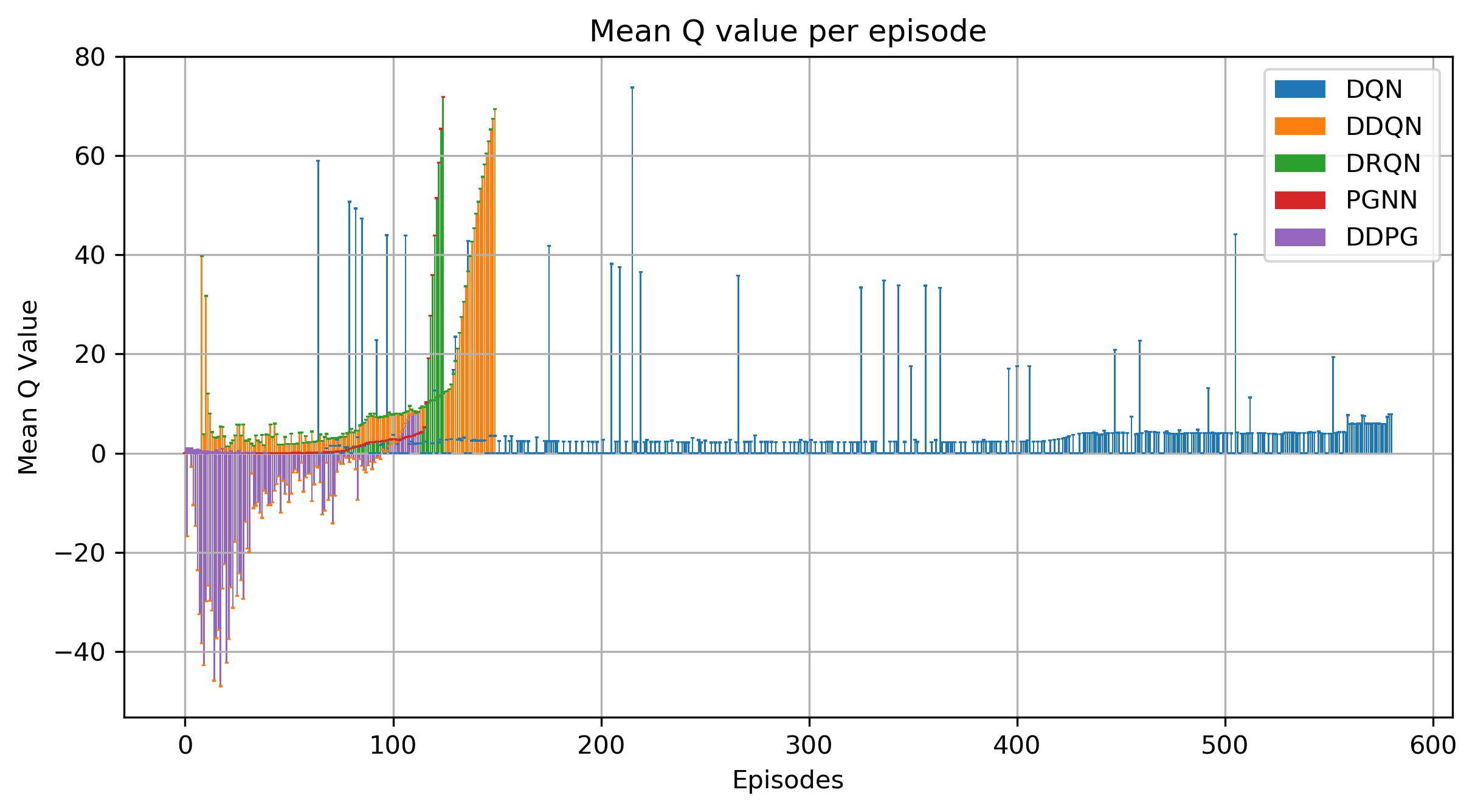}
\caption{Mean Q value per episode for (DQN, DDQN, DRQN, PGNN, and DDPG) }
\label{fig_mean_q}
\end{figure}

Figure \ref{fig_mean_q} depicts the  Mean of the Q-value for the five algorithms (DQN, DDQN, DRQN, PGNN, DDPG). From the figure it is clear that DRQN and DDQN both returned the heights Q-value. DRQN achieved the highest Q-value in 120 episodes, but DDQN achieved the highest Q-value in 155 episodes. The DDPG started by getting a huge negative reward and then it converged to positive value around 130 episodes as shown in Figure \ref{fig_mean_q}. The performance of PGNN was very poor as the mean Q-value was close to zero. However, the DQN performance was better than DDPG but it required a longer time to converge and yet it did not yield better Q-value after long training time.  

\begin{figure}
\centering
\includegraphics[scale=0.4]{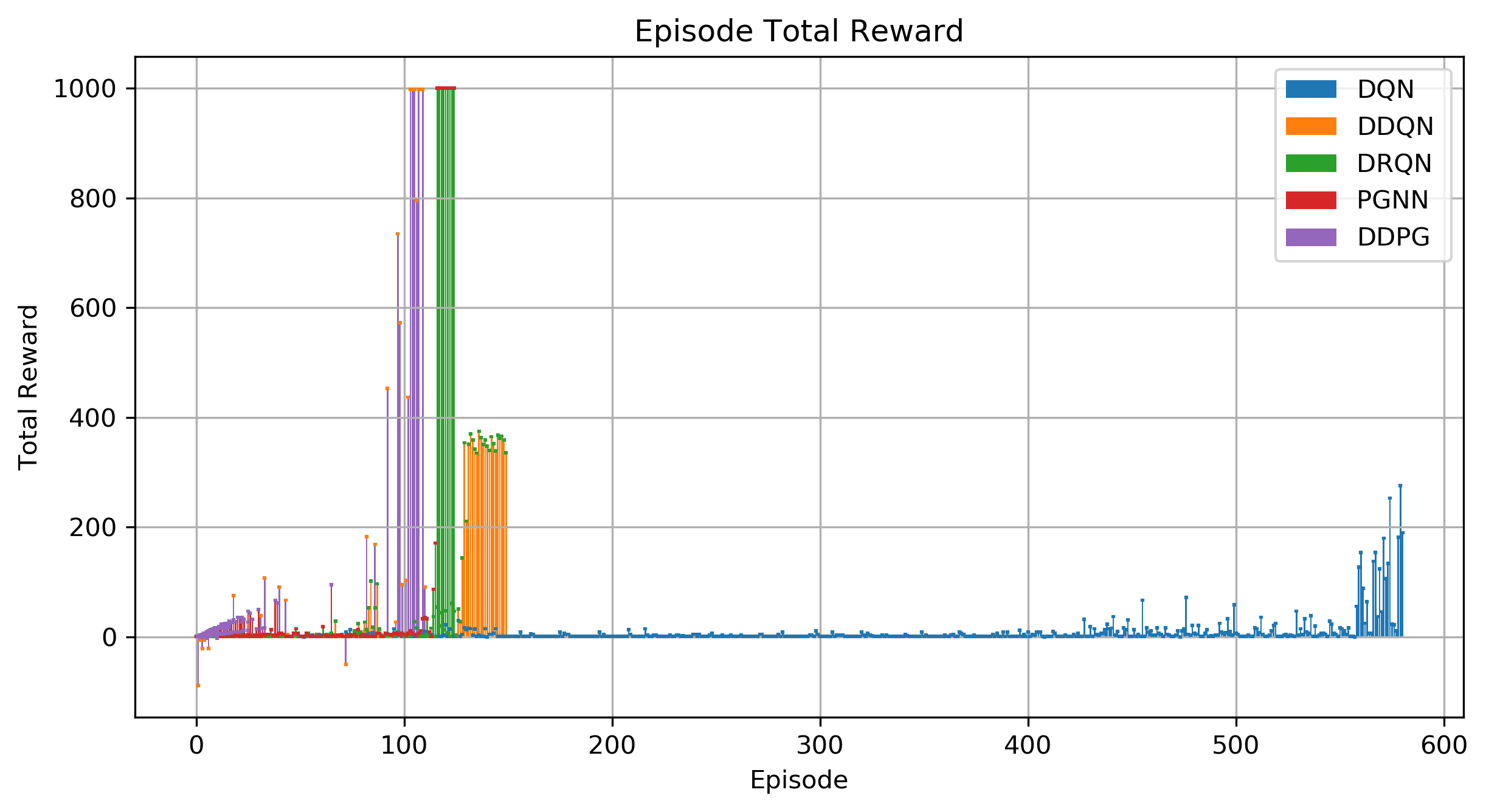}
\caption{Total Reward episode for (DQN, DDQN, DRQN, PGNN, and DDPG) }
\label{fig_total_reward}
\end{figure}

The divergence of the mean Q-values between the five algorithms can be justified by inspecting the total reward obtained by them, as shown in Figure \ref{fig_total_reward}. Looking at Figure \ref{fig_total_reward}, it was found that DRQN and DDPG managed to score the highest total rewards in 130 episodes. However, the DDQN came third in terms of the total obtained rewards, although DDQN takes longer time and more episodes of 600 to achieve the total rewards. However, the DQN outperform the PGNN in term of the total reward obtained but it takes the DQN the longest time to achieve this result (see Figure \ref{fig_total_reward}).

\begin{figure}
\centering
\includegraphics[scale=0.4]{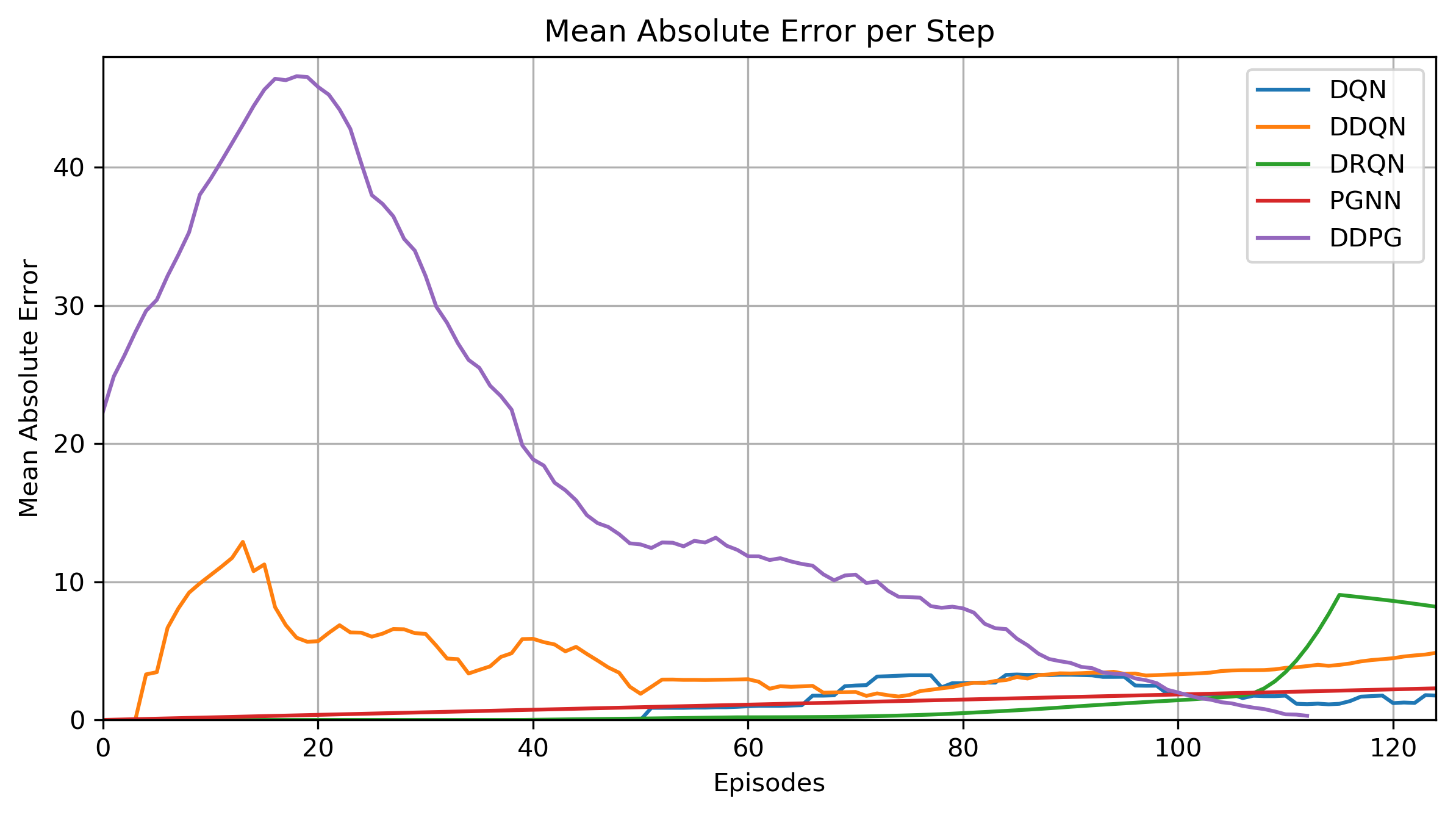}
\caption{Mean Absolute Error (MAE) for (DQN, DDQN, DRQN, PGNN, and DDPG)}
\label{fig_mae}
\end{figure}

\begin{figure}
\centering
\includegraphics[scale=0.4]{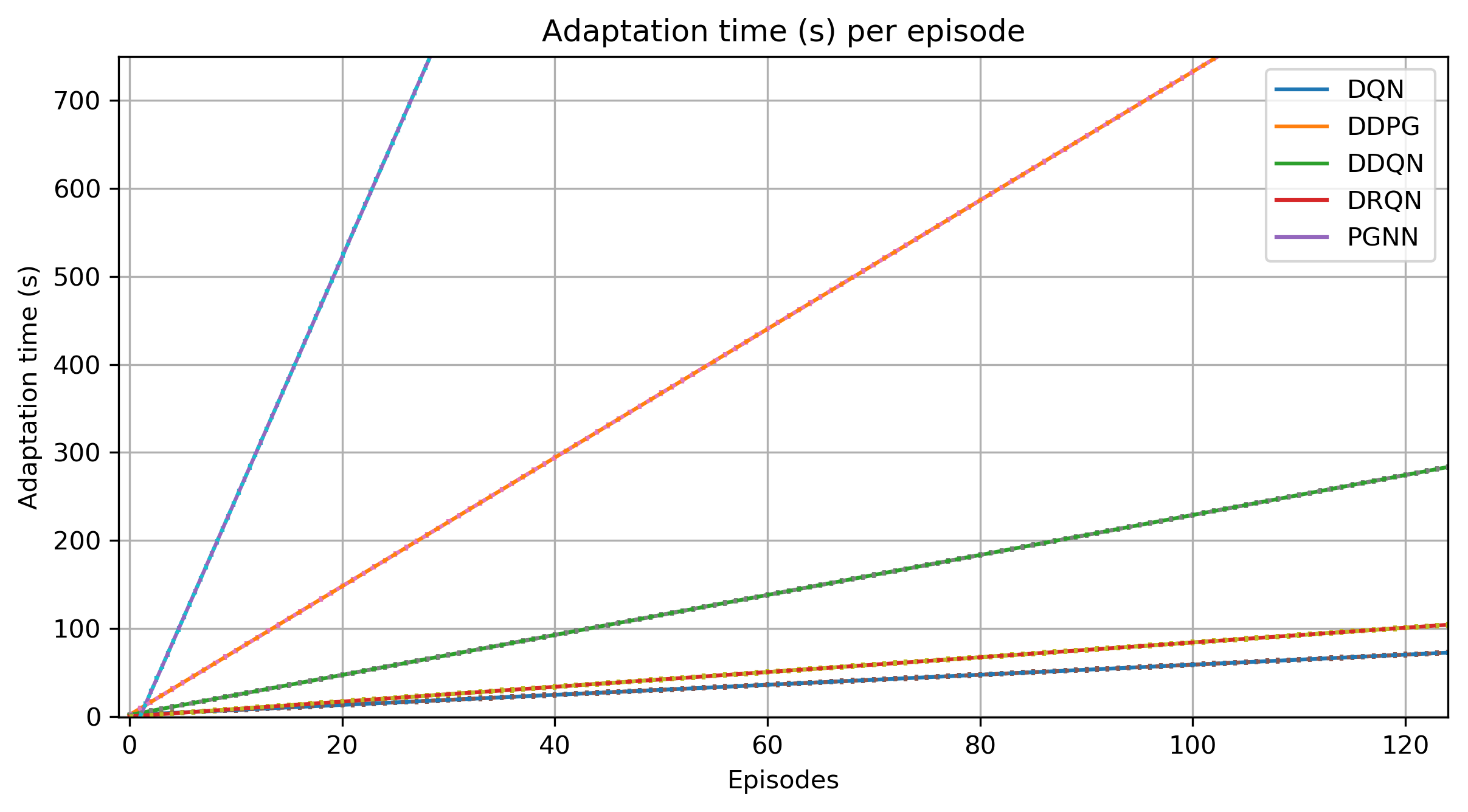}
\caption{Adaptation Time (s) Action Duration for  (MAE) for (DQN, DDQN, DRQN, PGNN, and DDPG)}
\label{fig_duration}
\end{figure}

On the other hand, the mean absolute error (MAE) of the five algorithms is illustrated in Figure \ref{fig_mae}. Figure  \ref{fig_mae} confirms the result of mean Q-value and total rewards described above. This figure shows that the DRQN, DDQN, and DQN algorithms produced fewer errors rate than the other two, as DRQN, DDQN, and DQN performed better in predicting the Q-value for each pair of state-action. The DDPG produces high errors rate and it failed to return an acceptable value of total rewards. This outcome is justified by how DDPG is working in the gradient ascent until it reaches a high negative reward value so it would work on the opposite direction - gradient descent - as this can be illustrated in Figure \ref{fig_mean_q}. However, the DQN produces the same MAE as DRQN and DDQN but it requires more training and episodes than the other four algorithms to achieve this performance. This can be explained by looking at the adaptation time/duration of the execution of all actions per episode in Figure \ref{fig_duration}. The DRQN, DDPG and DDQN  manage to converge faster and perform the adaptation in less time than DQN and PGNN. The PGNN requires longer time to complete the adaptation action, as it requires more time to adapt to the changes found in the observation. On the other hand, it was found that DRQN responded fastest to the changes in the observation space and managed to adapt quickly to the contextual changes. Also, the DQN and DDQN algorithm showed high levels of adaptation to changes but it takes both of them longer time to execute the adaptation with more number of episodes and steps. 

\begin{figure}
\centering
\includegraphics[scale=0.4]{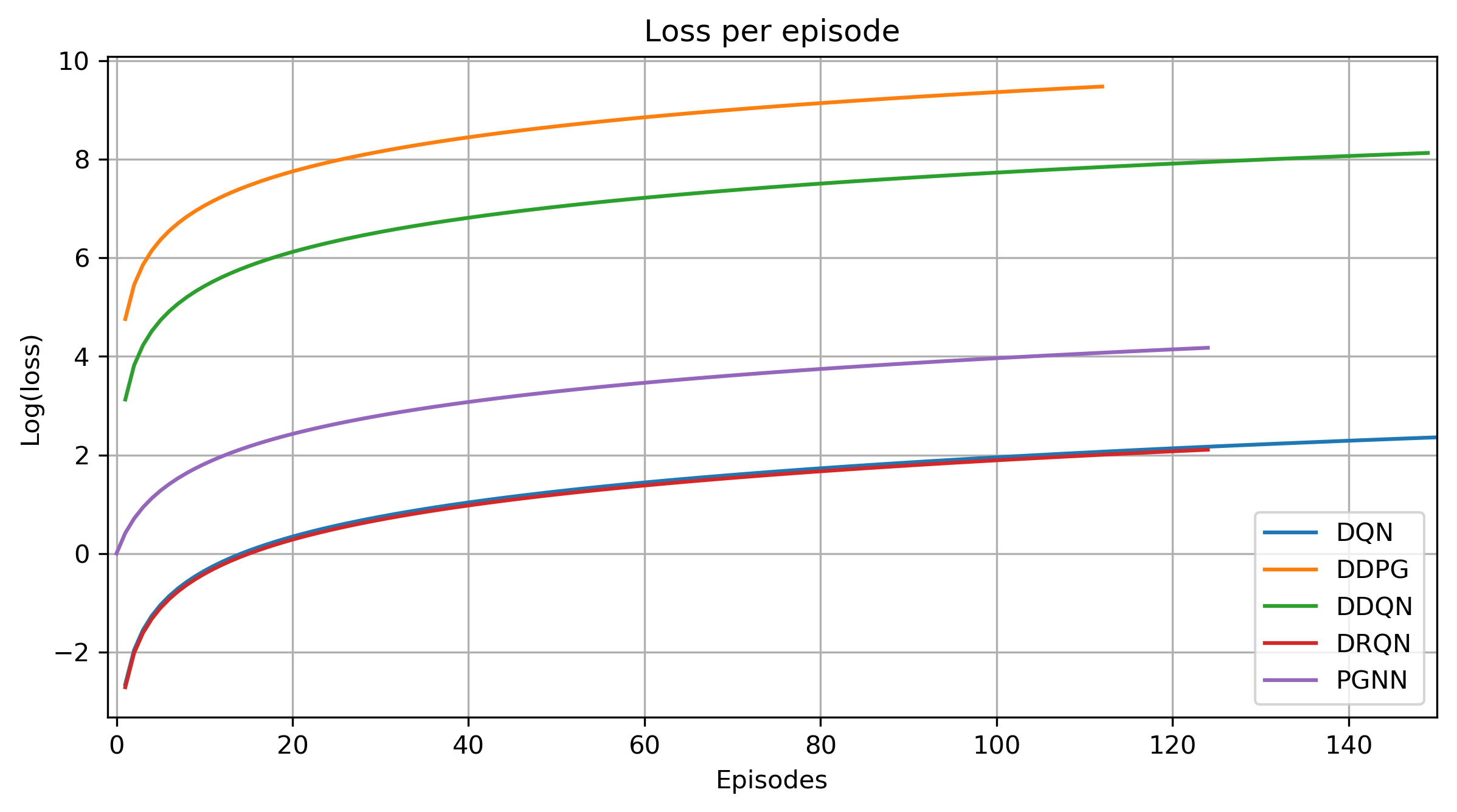}
\caption{Loss per episode for  (MAE) for (DQN, DDQN, DRQN, PGNN, and DDPG)}
\label{fig_loss}
\end{figure}

The excellent performance of DRQN in terms of the total rewards, maximum returned Q-value, less adaptation time and MAE is supported by the calculated loss for the five algorithms in Figure \ref{fig_loss}. The DDPG scores the highest loss, followed by PGNN. However, the DRQN comes fourth in term of loss. The DQN scores the best in terms of loss as the DQN algorithm takes a longer time to train, and the DQN collects more information about the environment, which results in a better loss that the other. However, the DQN has higher MAE as demonstrated in Figure \ref{fig_mae}. 

 \section{Conclusions and Future Work}
 \label{sec:Conclusion}
This paper presents i) a microservices architecture model that has continuous monitoring, continuous analysis of the observation space, and provides the architecture with dynamic decision making based on the employment of deep Q-learning/policy gradient algorithms. ii) MDP adaptation agents that can be used to observe the architecture's state spaces and executes an adaptation actions selected according to the outcome of the deep Q-learning/policy gradient algorithms. iii) An evaluation of five deep Q-learning/policy gradient algorithms including: deep Q-learning networks (DQN), duelling deep Q-learning networks (DDQN), deep recurrent Q-learning network (DRQN), policy gradient deep neural networks (PGNN), and deep deterministic policy gradient (DDPG). The evaluation in this research shows that DRQN is more suitable for driving the adaptation in POMDPs environment such as distributed microservices architecture. The DRQN shows a) faster training and convergence time, b) the highest maximum total reward in shortest number of episodes, c) faster adaptation time, and e) fewer errors rate and loss between the predicted and actual Q-value. 

This excellent performance of DRQN, when compared to the other four algorithms is justified by its ability:
 i) to reveal the patterns between the collected observations and the cumulative yielded rewards for each pair of state-action;
 ii) to perform back-propagation through time, which minimises the loss and error rate of the predicted Q-value;
 iii) to use GRU cells, so enabling the DRQN to possess a memory of all the yielded rewards from each pair of state-action pairs, which helps the DRQN to maximise the reward value quickly and reaches a optimal state faster than the other algorithms;
 iv) to overcome the problem of credit assignment and temporal-difference by stacking the collected observation instead of dealing with a single observation at a time, after which it uses its own memory to replay the last collected set of observation to the DRQN. This ability, in turn, gives the DRQN more insights about the rate of change, bias, and gradient decent of the observations.
The uses of Q-learning/policy gradient algorithms enable the architecture to dynamically elect a reasoning approach based on the highest reward gained from each action state pair. The self-adapting property is achieved by parameter tuning of the running services and dynamic composition/adjustment of the swarm cluster. We believe integrating reinforcement learning in the decision making process improves the effectiveness of the adaptation and reduces the adaptation risk, including the possibility of resources over-provisioning and thrashing. Also, our model preserves the cluster state by preventing multiple adaptations from taking place at the same time, as well as eliminating the actions that would return the lowest negative reward. Currently, this model can be extended by adding new actions to the action space implemented in the MDB agents, which will allow other researchers to run different types of experiments over, and informed by, this model. Our implementation is limited to a centralised multi-agents adaptation due to the limitation of a Docker swarm and the complexity to implement decentralised multi-agent RL algorithms. A Docker swarm enables the cluster to have one single leader, which prevents us from testing this model in multi agents’/leaders’ environments. Also, the current implementation of an adaptation agent is limited to discrete action space, which can be extended by adding more action policies to the action space of the MDP agent. Finally, we strongly believe in the ability of the service oriented architecture to reach high levels of self-adaptability by integrating MDP agents and deep recurrent Q-learning networks in a well designed MAPE-K architecture.

 \bibliographystyle{IEEEtran}
\bibliography{./selfhealingv2.bib} 

\begin{thebibliography}{10}
\providecommand{\url}[1]{#1}
\csname url@samestyle\endcsname
\providecommand{\newblock}{\relax}
\providecommand{\bibinfo}[2]{#2}
\providecommand{\BIBentrySTDinterwordspacing}{\spaceskip=0pt\relax}
\providecommand{\BIBentryALTinterwordstretchfactor}{4}
\providecommand{\BIBentryALTinterwordspacing}{\spaceskip=\fontdimen2\font plus
\BIBentryALTinterwordstretchfactor\fontdimen3\font minus
  \fontdimen4\font\relax}
\providecommand{\BIBforeignlanguage}[2]{{%
\expandafter\ifx\csname l@#1\endcsname\relax
\typeout{** WARNING: IEEEtran.bst: No hyphenation pattern has been}%
\typeout{** loaded for the language `#1'. Using the pattern for}%
\typeout{** the default language instead.}%
\else
\language=\csname l@#1\endcsname
\fi
#2}}
\providecommand{\BIBdecl}{\relax}
\BIBdecl

\bibitem{Oreizy:1999p3722}
P.~Oreizy, M.~M. Gorlick, R.~N. Taylor, D.~Heimhigner, G.~Johnson,
  N.~Medvidovic, A.~Quilici, D.~S. Rosenblum, and A.~L. Wolf, ``{An
  architecture-based approach to self-adaptive software},'' \emph{Intelligent
  Systems and Their Applications, IEEE}, vol.~14, no.~3, pp. 54--62, 1999.

\bibitem{Cheng:2009p3763}
B.~Cheng, R.~de~Lemos, H.~Giese, P.~Inverardi, J.~Magee, J.~Andersson,
  B.~Becker, N.~Bencomo, Y.~Brun, and B.~Cukic, ``{Software engineering for
  self-adaptive systems: A research roadmap},'' \emph{Software Engineering for
  Self-Adaptive Systems}, pp. 1--26, 2009.

\bibitem{bailey2011self}
C.~Bailey, D.~W. Chadwick, and R.~De~Lemos, ``Self-adaptive authorization
  framework for policy based rbac/abac models,'' in \emph{Dependable, Autonomic
  and Secure Computing (DASC), 2011 IEEE Ninth International Conference
  on}.\hskip 1em plus 0.5em minus 0.4em\relax IEEE, 2011, pp. 37--44.

\bibitem{Barbacci:2010p4077}
M.~Barbacci, ``{Software Quality Attributes: Modifiability and Usability},''
  \emph{Software Engineering Institute, Carnegie Mellon University, Pittsburgh
  PA}, vol. 15213.

\bibitem{van2012reinforcement}
H.~Van~Hasselt, ``{Reinforcement learning in continuous state and action
  spaces},'' in \emph{Reinforcement learning}.\hskip 1em plus 0.5em minus
  0.4em\relax Springer, 2012, pp. 207--251.

\bibitem{silver2016mastering}
D.~Silver, A.~Huang, C.~J. Maddison, A.~Guez, L.~Sifre, G.~Van Den~Driessche,
  J.~Schrittwieser, I.~Antonoglou, V.~Panneershelvam, M.~Lanctot, and {others},
  ``{Mastering the game of Go with deep neural networks and tree search},''
  \emph{Nature}, vol. 529, no. 7587, p. 484, 2016.

\bibitem{computing2006architectural}
A.~Computing \emph{et~al.}, ``An architectural blueprint for autonomic
  computing,'' \emph{IBM White Paper}, vol.~31, pp. 1--6, 2006.

\bibitem{ongaro2015raft}
D.~Ongaro and J.~K. Ousterhout, ``In search of an understandable consensus
  algorithm.'' in \emph{USENIX Annual Technical Conference}, 2014, pp.
  305--319.

\bibitem{bellman1957markovian}
R.~Bellman, ``{A Markovian decision process},'' \emph{Journal of Mathematics
  and Mechanics}, pp. 679--684, 1957.

\bibitem{wang2015dueling}
Z.~Wang, T.~Schaul, M.~Hessel, H.~Van~Hasselt, M.~Lanctot, and N.~De~Freitas,
  ``{Dueling network architectures for deep reinforcement learning},''
  \emph{arXiv preprint arXiv:1511.06581}, 2015.

\bibitem{hausknecht2015deep}
M.~Hausknecht and P.~Stone, ``{Deep recurrent q-learning for partially
  observable mdps},'' \emph{CoRR, abs/1507.06527}, vol.~7, no.~1, 2015.

\bibitem{sutton2000policy}
R.~S. Sutton, D.~A. McAllester, S.~P. Singh, and Y.~Mansour, ``Policy gradient
  methods for reinforcement learning with function approximation,'' in
  \emph{Advances in neural information processing systems}, 2000, pp.
  1057--1063.

\bibitem{silver2014deterministic}
D.~Silver, G.~Lever, N.~Heess, T.~Degris, D.~Wierstra, and M.~Riedmiller,
  ``{Deterministic policy gradient algorithms},'' in \emph{ICML}, 2014.

\bibitem{jelasityself}
O.~B.~M. Jelasity, A.~M.~C. Fetzer, S.~L.~A. van Moorsel, and M.~van Steen,
  ``Self-star properties in complex information systems,'' 2005.

\bibitem{horn:2001p3735}
P.~Horn, ``{Autonomic computing: IBM's Perspective on the State of Information
  Technology},'' Tech. Rep., 2001.

\bibitem{Cheng:2008p3708}
B.~Cheng, R.~de~Lemos, H.~Giese, P.~Inverardi, J.~Magee, R.~M. Malek,
  H.~M{\"u}ller, S.~Park, M.~Shaw, and M.~Tichy, ``{Software Engineering for
  Self-Adaptive Systems: A Research Road Map (Draft Version)},'' \emph{Dagstuhl
  Seminar Proc. 08031}, 2008.

\bibitem{Strang:2004p3770}
T.~Strang and C.~Linnhoff-Popien, ``A context modeling survey,'' in
  \emph{Workshop on advanced context modelling, reasoning and management,
  UbiComp}, vol.~4, 2004, pp. 34--41.

\bibitem{Cheng:2009p3902}
S.-W. Cheng, D.~Garlan, and B.~Schmerl, ``Evaluating the effectiveness of the
  rainbow self-adaptive system,'' in \emph{Software Engineering for Adaptive
  and Self-Managing Systems, 2009. SEAMS'09. ICSE Workshop on}.\hskip 1em plus
  0.5em minus 0.4em\relax IEEE, 2009, pp. 132--141.

\bibitem{MariusMikalsen:2005ur}
M.~Mikalsen, N.~Paspallis, J.~Floch, E.~Stav, G.~A. Papadopoulos, and
  A.~Chimaris, ``Distributed context management in a mobility and adaptation
  enabling middleware (madam),'' in \emph{Proceedings of the 2006 ACM symposium
  on Applied computing}.\hskip 1em plus 0.5em minus 0.4em\relax ACM, 2006, pp.
  733--734.

\bibitem{CheungFooWo:2007p1692}
D.~Cheung-Foo-Wo, J.~Y. Tigli, S.~Lavirotte, and M.~Riveill, ``{Self-adaptation
  of event-driven component-oriented middleware using aspects of assembly},''
  \emph{Proceedings of the 5th international workshop on Middleware for
  pervasive and ad-hoc computing: held at the ACM/IFIP/USENIX 8th International
  Middleware Conference}, pp. 31--36, 2007.

\bibitem{Wei:2016ge}
W.~Wei, X.~Fan, H.~Song, X.~Fan, and J.~Yang, ``{Imperfect Information Dynamic
  Stackelberg Game Based Resource Allocation Using Hidden Markov for Cloud
  Computing},'' \emph{IEEE Transactions on Services Computing}, vol.~11, no.~1,
  pp. 78--89, Feb. 2016.

\bibitem{Menasce:2007vq}
D.~A. Menasc{\'e} and V.~Dubey, ``Utility-based qos brokering in service
  oriented architectures,'' in \emph{Web Services, 2007. ICWS 2007. IEEE
  International Conference on}.\hskip 1em plus 0.5em minus 0.4em\relax IEEE,
  2007, pp. 422--430.

\bibitem{KonstantinosKakousis:2008ub}
K.~Kakousis, N.~Paspallis, and G.~A. Papadopoulos, ``Optimizing the utility
  function-based self-adaptive behavior of context-aware systems using user
  feedback,'' in \emph{OTM Confederated International Conferences" On the Move
  to Meaningful Internet Systems"}.\hskip 1em plus 0.5em minus 0.4em\relax
  Springer, 2008, pp. 657--674.

\bibitem{Sama:2008p3765}
M.~Sama, D.~S. Rosenblum, Z.~Wang, and S.~Elbaum, ``Model-based fault detection
  in context-aware adaptive applications,'' in \emph{Proceedings of the 16th
  ACM SIGSOFT International Symposium on Foundations of software
  engineering}.\hskip 1em plus 0.5em minus 0.4em\relax ACM, 2008, pp. 261--271.

\bibitem{Hirschfeld:2008p1620}
R.~Hirschfeld, P.~Costanza, and O.~M. Nierstrasz, ``Context-oriented
  programming,'' \emph{Journal of Object technology}, vol.~7, no.~3, pp.
  125--151, 2008.

\bibitem{Salehie:2009p3693}
M.~Salehie and L.~Tahvildari, ``{Self-adaptive software: Landscape and research
  challenges},'' \emph{Transactions on Autonomous and Adaptive Systems (TAAS},
  vol.~4, no.~2, May 2009.

\bibitem{RogeriodeLemos:2011tj}
R.~De~Lemos, H.~Giese, H.~A. M{\"u}ller, M.~Shaw, J.~Andersson, M.~Litoiu,
  B.~Schmerl, G.~Tamura, N.~M. Villegas, T.~Vogel \emph{et~al.}, ``Software
  engineering for self-adaptive systems: A second research roadmap,'' in
  \emph{Software Engineering for Self-Adaptive Systems II}.\hskip 1em plus
  0.5em minus 0.4em\relax Springer, 2013, pp. 1--32.

\bibitem{mnih2013playing}
V.~Mnih, K.~Kavukcuoglu, D.~Silver, A.~Graves, I.~Antonoglou, D.~Wierstra, and
  M.~Riedmiller, ``{Playing atari with deep reinforcement learning},''
  \emph{arXiv preprint arXiv:1312.5602}, 2013.

\bibitem{sutton1998introduction}
R.~S. Sutton and A.~G. Barto, \emph{{Introduction to reinforcement
  learning}}.\hskip 1em plus 0.5em minus 0.4em\relax MIT press Cambridge, 1998,
  vol. 135.

\bibitem{yoo2017action}
S.~Yoo, K.~Yun, J.~Y. Choi, K.~Yun, and J.~Choi, ``Action-decision networks for
  visual tracking with deep reinforcement learning.''\hskip 1em plus 0.5em
  minus 0.4em\relax CVPR, 2017.

\bibitem{caicedo2015active}
J.~C. Caicedo and S.~Lazebnik, ``Active object localization with deep
  reinforcement learning,'' in \emph{Proceedings of the IEEE International
  Conference on Computer Vision}, 2015, pp. 2488--2496.

\bibitem{jayaraman2016look}
D.~Jayaraman and K.~Grauman, ``Look-ahead before you leap: end-to-end active
  recognition by forecasting the effect of motion,'' in \emph{European
  Conference on Computer Vision}.\hskip 1em plus 0.5em minus 0.4em\relax
  Springer, 2016, pp. 489--505.

\bibitem{van2016deep}
H.~Van~Hasselt, A.~Guez, and D.~Silver, ``{Deep Reinforcement Learning with
  Double Q-Learning.}'' in \emph{AAAI}.\hskip 1em plus 0.5em minus 0.4em\relax
  Phoenix, AZ, 2016, p.~5.

\bibitem{hochreiter1997long}
S.~Hochreiter and J.~Schmidhuber, ``Long short-term memory,'' \emph{Neural
  computation}, vol.~9, no.~8, pp. 1735--1780, 1997.

\bibitem{williams1992simple}
R.~J. Williams, ``Simple statistical gradient-following algorithms for
  connectionist reinforcement learning,'' \emph{Machine learning}, vol.~8, no.
  3-4, pp. 229--256, 1992.

\bibitem{dowling2004self}
J.~Dowling and V.~Cahill, ``Self-managed decentralised systems using
  k-components and collaborative reinforcement learning,'' in \emph{Proceedings
  of the 1st ACM SIGSOFT workshop on Self-managed systems}.\hskip 1em plus
  0.5em minus 0.4em\relax ACM, 2004, pp. 39--43.

\bibitem{salehie2005policy}
M.~Salehie and L.~Tahvildari, ``A policy-based decision making approach for
  orchestrating autonomic elements,'' in \emph{Software Technology and
  Engineering Practice, 2005. 13th IEEE International Workshop on}.\hskip 1em
  plus 0.5em minus 0.4em\relax IEEE, 2005, pp. 173--181.

\bibitem{tesauro2007reinforcement}
G.~Tesauro, ``Reinforcement learning in autonomic computing: A manifesto and
  case studies,'' \emph{IEEE Internet Computing}, vol.~11, no.~1, 2007.

\bibitem{amoui2008adaptive}
M.~Amoui, M.~Salehie, S.~Mirarab, and L.~Tahvildari, ``Adaptive action
  selection in autonomic software using reinforcement learning,'' in
  \emph{Autonomic and Autonomous Systems, 2008. ICAS 2008. Fourth International
  Conference on}.\hskip 1em plus 0.5em minus 0.4em\relax IEEE, 2008, pp.
  175--181.

\bibitem{kim2009reinforcement}
D.~Kim and S.~Park, ``Reinforcement learning-based dynamic adaptation planning
  method for architecture-based self-managed software,'' in \emph{Software
  Engineering for Adaptive and Self-Managing Systems, 2009. SEAMS'09. ICSE
  Workshop on}.\hskip 1em plus 0.5em minus 0.4em\relax IEEE, 2009, pp. 76--85.

\bibitem{dutreilh2011using}
X.~Dutreilh, S.~Kirgizov, O.~Melekhova, J.~Malenfant, N.~Rivierre, and
  I.~Truck, ``Using reinforcement learning for autonomic resource allocation in
  clouds: towards a fully automated workflow,'' in \emph{ICAS 2011, The Seventh
  International Conference on Autonomic and Autonomous Systems}, 2011, pp.
  67--74.

\bibitem{jamshidi2016fuzzy}
P.~Jamshidi, A.~Sharifloo, C.~Pahl, H.~Arabnejad, A.~Metzger, and G.~Estrada,
  ``Fuzzy self-learning controllers for elasticity management in dynamic cloud
  architectures,'' in \emph{Quality of Software Architectures (QoSA), 2016 12th
  International ACM SIGSOFT Conference on}.\hskip 1em plus 0.5em minus
  0.4em\relax IEEE, 2016, pp. 70--79.

\bibitem{wu2018using}
T.~Wu, Q.~Li, L.~Wang, L.~He, and Y.~Li, ``Using reinforcement learning to
  handle the runtime uncertainties in self-adaptive software,'' in
  \emph{Federation of International Conferences on Software Technologies:
  Applications and Foundations}.\hskip 1em plus 0.5em minus 0.4em\relax
  Springer, 2018, pp. 387--393.

\bibitem{busoniu2008comprehensive}
L.~Busoniu, R.~Babuska, and B.~De~Schutter, ``A comprehensive survey of
  multiagent reinforcement learning,'' \emph{IEEE Transactions on Systems, Man,
  And Cybernetics-Part C: Applications and Reviews, 38 (2), 2008}, 2008.

\bibitem{marinescu2014decentralised}
A.~Marinescu, I.~Dusparic, A.~Taylor, V.~Cahill, and S.~Clarke, ``Decentralised
  multi-agent reinforcement learning for dynamic and uncertain environments,''
  \emph{arXiv preprint arXiv:1409.4561}, 2014.

\bibitem{lowe2017multi}
R.~Lowe, Y.~Wu, A.~Tamar, J.~Harb, O.~P. Abbeel, and I.~Mordatch, ``Multi-agent
  actor-critic for mixed cooperative-competitive environments,'' in
  \emph{Advances in Neural Information Processing Systems}, 2017, pp.
  6379--6390.

\bibitem{zhang2018fully}
K.~Zhang, Z.~Yang, H.~Liu, T.~Zhang, and T.~Ba{\c{s}}ar, ``Fully decentralized
  multi-agent reinforcement learning with networked agents,'' \emph{arXiv
  preprint arXiv:1802.08757}, 2018.

\bibitem{hernandez2018multiagent}
P.~Hernandez-Leal, B.~Kartal, and M.~E. Taylor, ``Is multiagent deep
  reinforcement learning the answer or the question? a brief survey,''
  \emph{arXiv preprint arXiv:1810.05587}, 2018.

\bibitem{mnih2016asynchronous}
V.~Mnih, A.~P. Badia, M.~Mirza, A.~Graves, T.~Lillicrap, T.~Harley, D.~Silver,
  and K.~Kavukcuoglu, ``Asynchronous methods for deep reinforcement learning,''
  in \emph{International conference on machine learning}, 2016, pp. 1928--1937.

\bibitem{kiss2017micado}
T.~Kiss, P.~Kacsuk, J.~Kovacs, B.~Rakoczi, A.~Hajnal, A.~Farkas, G.~Gesmier,
  and G.~Terstyanszky, ``Micado—microservice-based cloud application-level
  dynamic orchestrator,'' \emph{Future Generation Computer Systems}, 2017.

\bibitem{plappert2016kerasrl}
M.~Plappert, ``keras-rl,'' \url{https://github.com/keras-rl/keras-rl}, 2016.

\bibitem{nair2010rectified}
V.~Nair and G.~E. Hinton, ``Rectified linear units improve restricted boltzmann
  machines,'' in \emph{Proceedings of the 27th international conference on
  machine learning (ICML-10)}, 2010, pp. 807--814.

\bibitem{Bencomo:2010p3675}
N.~Bencomo, J.~Whittle, P.~Sawyer, A.~Finkelstein, and E.~Letier,
  ``{Requirements reflection: requirements as runtime entities},''
  \emph{Proceedings of the 32nd ACM/IEEE International Conference on Software
  Engineering-Volume 2}, pp. 199--202, 2010.

\bibitem{Medvidovic:1996p3905}
N.~Medvidovic, P.~Oreizy, J.~E. Robbins, and R.~N. Taylor, ``{Using
  object-oriented typing to support architectural design in the C2 style},''
  \emph{Proceedings of the 4th ACM SIGSOFT symposium on Foundations of software
  engineering}, pp. 24--32, 1996.

\bibitem{Paspallis:2009p3570}
N.~Paspallis, ``{Middleware-based development of context-aware applications
  with reusable components},'' \emph{University of Cyprus}, 2009.

\bibitem{monahan1982state}
G.~E. Monahan, ``State of the art—a survey of partially observable markov
  decision processes: theory, models, and algorithms,'' \emph{Management
  Science}, vol.~28, no.~1, pp. 1--16, 1982.

\bibitem{mou2017deep}
L.~Mou, P.~Ghamisi, and X.~X. Zhu, ``Deep recurrent neural networks for
  hyperspectral image classification,'' \emph{IEEE Trans. Geosci. Remote Sens},
  vol.~55, no.~7, pp. 3639--3655, 2017.

\bibitem{murad2017deep}
A.~Murad and J.-Y. Pyun, ``Deep recurrent neural networks for human activity
  recognition,'' \emph{Sensors}, vol.~17, no.~11, p. 2556, 2017.

\bibitem{cho2014properties}
K.~Cho, B.~Van~Merri{\"e}nboer, D.~Bahdanau, and Y.~Bengio, ``On the properties
  of neural machine translation: Encoder-decoder approaches,'' \emph{arXiv
  preprint arXiv:1409.1259}, 2014.

\bibitem{chung2014empirical}
J.~Chung, C.~Gulcehre, K.~Cho, and Y.~Bengio, ``Empirical evaluation of gated
  recurrent neural networks on sequence modeling,'' \emph{arXiv preprint
  arXiv:1412.3555}, 2014.

\bibitem{lillicrap2015continuous}
T.~P. Lillicrap, J.~J. Hunt, A.~Pritzel, N.~Heess, T.~Erez, Y.~Tassa,
  D.~Silver, and D.~Wierstra, ``Continuous control with deep reinforcement
  learning,'' \emph{arXiv preprint arXiv:1509.02971}, 2015.

\bibitem{daprato1995ornstein}
G.~Daprato and A.~Lunardi, ``On the ornstein-uhlenbeck operator in spaces of
  continuous functions,'' \emph{Journal of Functional Analysis}, vol. 131,
  no.~1, pp. 94--114, 1995.

\end{thebibliography}
\end{document}